\documentclass[twocolumn]{sigcomm-alternate}
\usepackage{graphicx,amsmath,amssymb,url,subfigure,epsfig,url,dcolumn,times}

\begin{document}

\title{The Workshop on Internet Topology (WIT) Report}

\author{
    Dmitri Krioukov \\ CAIDA \\ {\normalsize dima@caida.org} \and
    Fan Chung \\ UCSD \\ {\normalsize fan@math.ucsd.edu} \and
    kc claffy \\ CAIDA \\ {\normalsize kc@caida.org} \and
    Marina Fomenkov \\ CAIDA \\ {\normalsize marina@caida.org} \and
    Alessandro Vespignani \\ Indiana University \\ {\normalsize alexv@indiana.edu} \and
    Walter Willinger  \\ AT\&T Research \\ {\normalsize walter@research.att.com}
}

\maketitle
\date{}

\pagestyle{empty}
\thispagestyle{empty}

\begin{abstract}
\noindent Internet topology analysis has recently experienced a
surge of interest in computer science, physics, and the mathematical
sciences. However, researchers from these different disciplines
tend to approach the same problem from different angles.
As a result, the field of Internet topology analysis and modeling must
untangle sets of inconsistent findings, conflicting claims, and
contradicting statements.

On May 10-12, 2006, CAIDA hosted the Workshop on Internet topology~(WIT).
By bringing together a group of researchers spanning the areas of computer
science, physics, and the mathematical sciences, the workshop aimed to
improve communication across these scientific disciplines, enable
interdisciplinary cross-fertilization, identify commonalities in the
different approaches, promote synergy where it exists, and utilize the
richness that results from exploring similar problems from multiple
perspectives.

This report describes the findings of the workshop,
outlines a set of relevant open research problems identified by
participants, and concludes with recommendations
that can benefit all scientific communities interested
in Internet topology research.
\end{abstract}




\section{Key findings}
\label{sec:workshop}

{\bf Motivation.} Different communities study the Internet topology
from different perspectives and for different reasons.

To {\em networking researchers}, the term ``Internet topology'' is
multi-faceted, and the precise meaning depends on what a node or a link
represents, which in turn can differ across different layers of the
Internet architecture, e.g., physically meaningful topologies such as the
router-level connectivity, or more logical constructs such as AS-level topology,
or overlay networks such as the WWW graph, email graph, P2P networks.
The networking research motivation for studying Internet-specific
topologies is to enable prediction of how new technologies, policies, or
economic conditions will impact the Internet's connectivity structure at
different layers.

To {\em non-networking researchers}, and especially to {\em
physicists}, the Internet is just one of many examples of a complex
network, albeit one uniquely amenable to measurements and
experimentation because it is man-made. Their motivation for
studying Internet topology is generally more fundamental than that of
networking researchers. Physicists search for inherent principles
shaping small- and large-scale network patterns. They want to find
universal laws of the evolution of complex systems that transcend
specific application domains.

{\em Mathematicians\/} do not necessarily seek connections between
their purely abstract theories and the real world. But the other
communities recognize the need for a rigorous framework to support
Internet topology analysis, and hope that having mathematicians involved
will stimulate the development of suitable mathematical apparatuses.

{\em Engineers\/} need to better understand the Internet structure
since performance of several applications and protocols depends
strongly on peculiarities of an underlying network. For example,
there is a proven huge gap between the best possible performance of
routing on random graphs and on trees or
grids~\cite{gavoille01,ThoZwi01b}. Recent research suggests that
observed Internet-like topologies are particulary well-structured
for routing efficiency~\cite{KrFaYa04,BaCo06}, but the existing
Internet routing architecture does not exploit this efficiency. The
knowledge and understanding of the topological properties of the
Internet should help engineers to optimize future technological
developments.

Despite the diverse motivations described above, researchers from
different disciplines all agree that we need to identify and
understand the essential properties that are responsible for certain
behaviors of certain applications. {\em Predictive power\/} is
therefore regarded by all communities as the Holy Grail
of Internet topology research, cf.~\cite{nas-report}.

\vskip 6pt {\bf Models.} There are numerous models of the Internet
topology. We can roughly distinguish them as static, i.e.,
constructing statistical ensembles of random networks with certain
characteristics matching values measured in the real Internet, and
dynamic, i.e., trying to reproduce the details of the Internet
evolution/growth.
The models of the former type tend to be
descriptive, while the
models of the latter type can be explanatory.

Another dimension in model classification encompasses a trade-off between:
1)~complexity of a model and the amount of observable details it tries to
reproduce, and 2)~its explanatory power and associated generality.
At one extreme are models striving to blindly reproduce all the
details of the observed complex phenomenon, e.g., the Internet.
These approaches usually includes numerous assumptions and a huge
number of parameters that often make the model not
transparent and with a low explanatory or predictive power.
At the other extreme are
``conceptual models'' that might have an appealing
theoretical value promising the most fundamental insights of general
nature, but that reproduce no specific characteristic of a given
system and thus have no practical applications or predictive power
either. Finding the right
balance between these two extremes is of critical importance
to understanding complex systems, in general, and the Internet,
in particular.

{\em Networking researchers\/} increasingly look for and demand
network models that are not only descriptive in the sense of
matching certain graph-theoretic properties, but that also have
network-intrinsic meaning, provide context for known structural or
architectural features of Internet, and withstand scrutiny against data
and by domain experts.

To {\em physicists}, the insistence on specificity and pursuit of
models reflecting networking reality has to be carefully balanced
since the profusion of constraints tends to rule out more general modeling
approaches where abstraction and generality are key elements usually
hindered by the inclusion of specialized
design features~\cite{PaSaVe04-book}.

One of the essential differences between the approaches of
these two communities to modeling and explaining Internet-related
topologies is the role of randomness.
The desire for abstraction and resilience to system-specific details
renders randomness a critical component in physics-inspired models.
An example is the preferential attachment toy model~\cite{BarAlb99},
where the network emerges as a result of the contrast between the
randomness and the preference function, as encoded
in the form of the attachment probability. In contrast, randomness
plays a relatively small role in the ``first-principles'' approach to
Internet router-level topology modeling exemplified by the
heuristically optimized tradeoff (HOT) toy model~\cite{LiAlWiDo04}.
In this model, randomness enters only with the purpose of
accounting for uncertainties in the environment, e.g., traffic demands,
while the core of the model derives from deterministic design decisions
that seek to optimize certain domain-specific and technological
network characteristics.
These two models are both capable of accounting for
the high variability in node degree distributions,
but they otherwise starkly differ, in terms of generation,
evolution, and structural properties.

A path to common ground is finding
interdependencies between metrics employed to generate and
characterize network topologies~\cite{MaKrFaVa06}. As soon as two
different topology characteristics are found to be related, any two
models based on these two different metrics are necessarily allied
as well, even if they originate as completely different or even
mutually exclusive. Consider the HOT-inspired FKP model
in~\cite{FaKoPa02} that was originally envisioned as having nothing
in common with preferential attachment. One of the trade-off
optimization objectives in the FKP model is minimization of the
average distance from the attachment node to the rest of the
network. Since this distance directly depends on the degree of the
node~\cite{DoMeOl06,HoSiFro05}, the model actually reduces to
a form of the preferential attachment model, albeit with no power
laws~\cite{AlDo04,BeBoBo03}. Analogously, the introduction of more
complicated and constrained generating rules in stochastic evolving
networks may effectively account for design principles of increasing
complexity that often compete among themselves, leading to a convergence
of modeling perspectives~\cite{DorMen-book03,PaSaVe04-book}.

In other words, interdependencies between different metrics can identify
and explain similarities among low-order approximations of various
complex systems, e.g., their representative graphs. At the same
time, higher order detail of the correlation functions
characterize the differences among these systems.
Indeed, the finer the granularity we use to describe networks,
the more differences (and noise) we must expect to see
among different instantiations.

\vskip 6pt
{\bf Data.} True predictive models of the Internet topology and evolution
cannot be developed without validation by real data. In its current state,
Internet topology research is not an informed discipline since
available data is not only scarce, but also severely limited by technical,
legal, and social constraints on its collection and distribution.

Different communities may have different views of and needs for the
data. {\em Mathematicians} do not need data at all. {\em Physicists}
are interested in data to support their models, but are not especially
concerned much about the data quality. They tend to
take available data at face value and disregard domain-specific
details as statistically insignificant. Both these communities have
to rely upon the expertise of the networking community in selecting
the most reliable and suitable data for analysis.

{\em Networking researchers} have come to realize the limitations, ambiguities,
and shortcomings of the measurements that form the basis of existing
Internet topology research. In fact, there has been an increasing awareness
that much of the available data cannot and should not be used at face value.
Demonstrating the robustness of an inferred property to
the most glaring ambiguities in the data sets is as important (if not more)
as establishing the property in the first place.

{\em Engineers} are the closest to collecting actual data, at least
about their own networks. However, data ownership and stewardship are complex
and highly charged issues with numerous social, political, liability,
and security implications. As was recently demonstrated by the AOL fiasco
with publishing anonymized search results~\cite{aol-data}, commercial and legal
pressures render it close to impossible to channel Internet measurement data
from private enterprises to the research community.

All communities agree that a lack of comprehensive high-quality topological
and traffic data is highly detrimental to the progress of  Internet
infrastructure research, cf.~\cite{nas-report}. A constant push for access to more and better
data requires concerted efforts from all communities involved.

At the same time, it is clear that most of the measurement-related problems
will not disappear soon and that future topology and traffic data will always
be of somewhat limited quality. It is the responsibility of the networking
community to point out assumptions and limitations of measurement
experiments and explain the ambiguities in the resulting data.
It is the responsibility of all data users to educate themselves on
the incompleteness, inaccuracy, and other deficiencies of these
measurements and to avoid overinterpretation.

\vskip 6pt {\bf Outreach.} The current bottleneck remains
interdisciplinary communication, cf.~\cite{nas-report}. Although the different communities
generally agree on the research objectives, formalizations of
problems are often so drastically different that it is hard to
understand each other or see common ground. Each community feels
that the others need to be more receptive to and able to use
insights that derive from looking at similar types of problems in a
number of different ways.

Unfortunately, non-networking researchers sometimes have problems with
publishing their work in networking journals, conferences, or workshops.
Some have noted that the reviewers are overly concerned with domain-specific
details and pay little or no attention to the potential novelty of approaches
employed by other disciplines. At the same time, networking researchers
expect papers submitted to networking journals and conferences to include
an appropriate networking context for abstract or more graph-theoretic
work, along with an illustration of how the results in the paper
provide new acumen for networking.

To increase the bandwidth and efficacy of the dialogue among the different
communities, CAIDA held the first Workshop on Internet Topology~\cite{wit}.
Of the roughly 40 invited participants, about 30\% represented the
physical sciences, 60\% computer science/engineering, and 10\% the
mathematical sciences. Almost 50\% of the participants were graduate
students or postdocs working on Internet topology-related problems.
Lively engagement of representatives from different disciplines contributed
to the success of WIT in facilitating a productive exchange of ideas and
arguments.

The workshop started with two tutorial-style talks.
Alessandro Vespignani first gave a careful introduction
to Internet modeling from the physics perspective. He was followed by
David Alderson, who illustrated the networking perspective by focusing on
modeling the Internet's router-level topology. A number of
presentations addressed problems with Internet topology
measurements, including incomplete and inaccurate data due to
statistical sampling biases and/or an inability to detect and identify
connectivity below the IP layer. Another set of talks dealt with
different approaches to Internet topology modeling and provided
examples of descriptive vs.\ explanatory models and equilibrium vs.\
non-equilibrium models. A number of talks treated the Internet as a
correlated network, and problems of interest included extracting and
understanding the underlying correlation structure, studying the
interdependencies among different network properties, and exploring
the diversity within the space of certain classes of correlated
network models. The workshop concluded with a half-day of
discussions, and the following sections provide a summary of the
open research problems and recommendations that were identified and
articulated during these discussions.

For detailed information about the meeting presentations, please see
the meeting agenda~\cite{wit} with links to the actual slides in the
PDF format.

\section{Open problems}
\label{sec:problems}

\subsection{Data}
\label{sec:data:problems}
Researchers recognize that despite their limitations, the available
measurements do
provide valuable information, and the challenge is to extract that information
and use it in an appropriate and adequate manner. The WIT participants
acknowledged the need for better Internet topology data and for better access
to existing data, cf.~\cite{nas-report}, and identified the following unresolved problems.
\begin{enumerate}
\item
All measurements are constrained by experimental and observational
conditions, i.e., lack of observation points, finite number of
destinations probed, inability to capture other layers and
disambiguate between high-degree nodes and opaque clouds, etc., and
as a result, produce incomplete, inaccurate, and ambiguous data. We
need to optimize our data collection and validation efforts, and to
develop methods for objective assessment of measurement quality.

\item Incompleteness of the data may distort our view of the Internet by
causing biases in derived topologies at the router- or the AS-level.
The probability of strong, qualitative differences between reality
and observations is low: it was shown that specific graphs
classes, e.g., classical Erd\H{o}s-R\'{e}nyi random graphs, are
extremely unlikely to represent real Internet topologies measured from
multiple vantage points~\cite{DaAlHaBaVaVe05}. At the same time,
inference of probability distributions specifying possible
quantitative deviations of real topologies from measured ones
remains largely an open problem, even though there have been some
recent attempts to address it~\cite{ViBaDaZhKo05}.

\item We need targeted measurements focused on particular geographic areas. By
comparing and contrasting data from different geopolitical and socioeconomic
environments researchers will distinguish between global core properties of the
Internet and its locally specific manifestations.

\item Internet measurement would ideally progress from measuring only the
intra- and inter-AS topology at the router- and AS-level to measuring
link bandwidths and actual traffic flows on a representative portion of
the Internet, cf.~\cite{nas-report}. These tasks are notoriously difficult: even proposing
and implementing novel kinds of measurements is a challenging task,
and existing measurement tools have not demonstrated the ability to
scale up to measure link and/or node properties across realistic networks.
Furthermore, making progress in this area is unlikely without protected
access to the infrastructure components that need to be measured.
For recent attempts to address these problems, see~\cite{ChaJaMa05}.

\end{enumerate}

\subsection{Modeling}
We characterize and model the Internet via different formalisms and at
different levels of abstraction. We recognize that all models are imperfect
and incomplete, and scientific progress often requires having a more than
one model for the same phenomenon.  The following specific problems were
discussed at the workshop.

\begin{enumerate}
\item {\em Descriptive models} strive to reproduce some graph-theoretic
properties of the Internet and usually are not concerned with
their network-specific interpretation. A review
relating graph-theoretic parameters to corresponding practically important
network characteristics in~\cite{MaKrFo06} offers a modest beginning toward
bridging this gap.
In contrast, {\em explanatory models} typically acknowledge and respect
domain-specific constraints while attempting to simulate the
fundamental principles and factors responsible for the
structure and evolution of
network topology, e.g., traffic conditions, cost-minimization requirements,
technological reality. Yet determining which forces and factors are
critical to faithful modeling of Internet topology and evolution
is a glaring open problem.

\item One of the less intuitively satisfying approaches to model
fitting is to match an increasing number of graph metrics with
corresponding statistics of inferred Internet connectivity.
This exercise can be interminable, and yields little insight
into essential properties of networks. The matching exercise also does
not constitute a sufficient model validation, especially in view
of the limited quality of the available measurements.
There was consensus  at the workshop for proper comparison and
validation methodologies.\\
\hspace*{4pt} (i) Not all topology metrics are mutually independent:
some either fully define others or, at least, significantly narrow
down the spectrum of their possible values. Therefore, identifying
bases of such definitive metrics reduces the number of topology
characteristics that explanatory models must reproduce. The
$dK$-series~\cite{MaKrFaVa06} presents one possible approach to
constructing a family of such simple metrics defining all others.
Are there other bases, different from the $dK$-series, that
carry the same properties?\\
\hspace*{4pt} (ii) The desired accuracy in matching various
topological parameters should depend on the question posed.
For example, if the performance of a routing algorithm
depends only on the distance distribution in the network, then two
topologies match perfectly as soon as their distance
distributions are the same, independent of other characteristics.\\
\hspace*{4pt} (iii) All models should be based on {\em physical},
that is, {\em measurable\/} external parameters. Many
non-physical parameters employed in a model explode the
exploration space, allowing one to freely tune these unmeasurable
parameters to match the model output with empirical data. But
this approach by definition denies the possibility of true validation
of the model which degrades its conceptual value. Such non-physical models
should be assiduously avoided, or at least they must include suggestions
on how to measure/validate values of their most crucial external parameters.

\item Future developments in the field of Internet modeling may include the
following advancements, although we recognize the unlikelihood of achieving
these goals without support of infrastructure owners:\\
\hspace*{4pt} (i)
annotated models of an ISP's router-level topology, where nodes are labeled
with router capacity, type, or role, and link labels describe delay, distance,
or bandwidth;\\
\hspace*{4pt} (ii)
annotated models of the Internet's AS-level topology, where node labels
include AS-specific information, e.g., number and/or locations of PoPs,
customer base, and link labels reflect peering relationships;\\
\hspace*{4pt} (iii)
models built around parameters closely related to real use of the network,
e.g., routing models that define and utilize routing-related parameters such
as robustness, fairness, outage, etc.;\\
\hspace*{4pt} (iv)
dynamic, evolutionary models of the Internet deriving simple rules for network
evolution from actual technological constraints, e.g., from known
Cisco router characteristics.
\end{enumerate}

\subsection{General Theory}

At the AS level, the Internet topology is a result of local business
decisions independently made by each AS. Since there is no explicit global
human control or design of the AS-level topology, it is often considered as an
example of a self-evolved and self-organized system. On the other hand, at the
router level the Internet topology is a product of human-controlled
technological optimizations aiming to minimize cost and maximize
efficiency. The presence of such elements of design and engineering
makes the Internet a complex {\em engineered\/} system.

Specific theoretical topics discussed at the workshop included:
\begin{enumerate}
\item So far, graph theory has provided the mathematical apparatus most
commonly used for network research. Is traditional graph theory suitable for
dealing with dynamic network structures that change over
time?  Is it even the right underlying theory for network
structure in face of mobility, delay-tolerant networks, and other
technological advances?

\item Multiple layers in the Internet protocol stack have their own
corresponding topologies, i.e., fiber, optical, router, AS, Web, P2P
graphs, that describe significantly different aspects of Internet
connectivity. The
challenge is to develop a proper mathematical framework that would
provide an efficient and accurate mapping between such different
descriptions while retaining the network-specific meaning at the
various levels of abstraction. Multiscale analysis, modeling, and
simulation~\cite{mms04-book,mms-journal}, done in a coherent manner,
seem promising for dealing
with the multiscale nature of Internet connectivity and dynamics of
heterogeneous, and potentially annotated, layer-specific structures.

\item We cannot effectively explain Internet-related topologies
without a basic understanding of the traffic exchanged across
these connectivity structures, e.g., AS-level traffic matrices~\cite{ChaJaMa05},
cf.~\cite{nas-report}.
As described in Section~\ref{sec:data:problems}, data in support of
this kind of correlation is extremely limited at present, but
the needs articulated by theorists may eventually become a driving
force stimulating development of new approaches, techniques, and
tools for measuring, or at least inferring, AS-related traffic
quantities.

\item It is unclear how the interplay among economical, political, social
forces, on one hand, and technological realities, on the other hand,
shapes the past, present, and the future of the Internet. For
example, is the router-level topology of a large Korean ISP different
because of their atypically high penetration of broadband deployment,
or importance of gaming traffic? A recent study~\cite{ZhZhZh06} claims
that the (still relatively) small Chinese Internet AS-level topology
preserves the structural
characteristics of the global Internet and follows the same
evolution dynamics despite being developed with more centralized
planning and less commercial competition. If correct, such results would
emphasize the primary role of technological factors, such as performance
metrics and equipment constraints, which are fairly universal across
the globe. Understanding of a sociopolitical foundation of the
observed Internet topology remains an elusive goal and further
research aimed at its quantitative characterization should be
supported.
\end{enumerate}

\section{Recommendations}
\label{sec:recom}

{\bf Interdisciplinary communication remains a serious bottleneck.}
The science of the Internet is multidisciplinary and requires continual
cross-fertilization among networking, physics, mathematics, and
engineering communities. Each community should increase its openness
to results from other communities. It is extremely important to
read, try to understand, and cite publications from other fields. To
facilitate the interdisciplinary flow of knowledge we
recommend the following steps:\\
\hspace*{4pt} (i)
regular interdisciplinary meetings that target researchers from
specific scientific communities and enable the exchange of ideas and
demonstration of new approaches;\\
\hspace*{4pt} (ii)
educational outreach by offering more interdisciplinary classes,
developing interdisciplinary tutorials, vocabularies, educational web pages
that foster the exchange of relevant domain knowledge;\\
\hspace*{4pt} (iii)
student involvement at early stages so they grow familiar with the
literature in the different fields and can become ``bridge-builders''
among the different groups.

\vskip 6pt
{\bf A lack of comprehensive and high-quality topological and traffic data
represents a serious obstacle to successful Internet topology modeling, and
especially model validation.} To improve the current situation we
recommend:\\
\hspace*{4pt} (i)
outreach to Internet registries, e.g., ARIN, RIPE, and other databases
regarding access and use of their data for research purposes;\\
\hspace*{4pt} (ii)
develop new techniques and tools to collect the data for the next
generation of Internet models;\\
\hspace*{4pt} (iii)
encourage researchers to use the data to account for known
deficiencies in their analysis and to demonstrate that obtained results
are robust;\\
\hspace*{4pt} (iv)
support repositories of publicly available topology and traffic data
that clearly identify limitations and shortcomings of the data.

Official repositories of publicly available data exist in many
``data-intensive'' sciences. A good example is the Protein Data
Bank~\cite{pdb} in chemistry. Newly discovered proteins must be
indexed there before papers referring to them can be published.

We note that in June 2006, one month after WIT, CAIDA opened for public
browsing the catalog of Internet measurement data, DatCat~\cite{datcat}.
The main goal of DatCat is to facilitate sharing of data
sets with researchers in pursuit of more reproducible scientific results.
Connecting researchers to available datasets will maximize the research
use of existing Internet data and hopefully promote a stronger requirement
for validation in the field ~\cite{datcat-ccr}.
As of October 2006, the catalog indexed 4.8 TB of CAIDA data. We
are working with selected owners of other Internet data collections to
help them index their data into DatCat. We are also working on a public
contribution interface that would allow anyone in the community to index
their datasets in the catalog.

One of the core features of the DatCat that directly addresses a need
articulated at WIT is the ability for users to add annotations to catalog
objects. By annotating data, investigators with experience in analyzing a
particular dataset will be able to share with others their important
findings including key statistics, novel features, bugs, caveats, and
any other relevant information about a given dataset.

\vskip 6pt
{\bf The networking research community must do better at promoting
Internet topology research, both its scientific merit and its broader impact.}
Our suggestions include:\\
\hspace*{4pt} (i)
endeavor to convert theoretical results into practical solutions
that matter for real networks, e.g., performance, revenue,
engineering, etc.;\\
\hspace*{4pt} (ii)
make exchange of information and ideas between scientists
and engineers a priority;\\
\hspace*{4pt} (iii)
work with funding and science policy agencies to disseminate and
implement the ideas and recommendations from this workshop.

In particular, the design plans for the Global Environment for Network
Innovations (GENI)~\cite{geni} currently under consideration at the NSF
is a potential area of impact.  Can a GENI-like facility help
in tackling some of the research challenges identified in this report,
and if so, how?

\vskip 6pt
{\small ACKNOWLEDGMENTS.} The workshop was supported by NSF grant CNS-0434996,
CAIDA members and the San Diego Supercomputer Center. We thank all the
participants for their insights and support.

\bibliographystyle{IEEE}
\bibliography{bib}

\begin{thebibliography}{10}

\bibitem{gavoille01}
C.~Gavoille,
\newblock ``Routing in distributed networks: Overview and open problems,''
\newblock {\em ACM SIGACT News - Distributed Computing Column}, vol. 32, no. 1,
  pp. 36--52, 2001.

\bibitem{ThoZwi01b}
M.~Thorup and U.~Zwick,
\newblock ``Compact routing schemes,''
\newblock in {\em SPAA}, 2001.

\bibitem{KrFaYa04}
D.~Krioukov, K.~Fall, and X.~Yang,
\newblock ``Compact routing on {Internet}-like graphs,''
\newblock in {\em INFOCOM}, 2004.

\bibitem{BaCo06}
A.~Brady and L.~Cowen,
\newblock ``Compact routing on power-law graphs with additive stretch,''
\newblock in {\em ALENEX}, 2006.

\bibitem{nas-report}
{\em Network Science},
\newblock The National Academies Press, Washington, 2006.

\bibitem{PaSaVe04-book}
R.~Pastor-Satorras and A.~Vespignani,
\newblock {\em Evolution and Structure of the {Internet}: A Statistical Physics
  Approach},
\newblock Cambridge University Press, Cambridge, 2004.

\bibitem{BarAlb99}
A.-L. Barab\'{a}si and R.~Albert,
\newblock ``Emergence of scaling in random networks,''
\newblock {\em Science}, vol. 286, pp. 509--512, 1999.

\bibitem{LiAlWiDo04}
L.~Li, D.~Alderson, W.~Willinger, and J.~Doyle,
\newblock ``A first-principles approach to understanding the {Internet's}
  router-level topology,''
\newblock in {\em SIGCOMM}, 2004.

\bibitem{MaKrFaVa06}
P.~Mahadevan, D.~Krioukov, K.~Fall, and A.~Vahdat,
\newblock ``Systematic topology analysis and generation using degree
  correlations,''
\newblock in {\em SIGCOMM}, 2006.

\bibitem{FaKoPa02}
A.~Fabrikant, E.~Koutsoupias, and C.~H. Papadimitriou,
\newblock ``Heuristically optimized trade-offs: A new paradigm for power laws
  in the {Internet},''
\newblock in {\em {ICALP}}, 2002.

\bibitem{DoMeOl06}
S.~N. Dorogovtsev, J.~F.~F. Mendes, and J.~G. Oliveira,
\newblock ``Degree-dependent intervertex separation in complex networks,''
\newblock {\em Physical Review E}, vol. 73, pp. 056122, 2006.

\bibitem{HoSiFro05}
J.~A. Ho{\l}yst, J.~Sienkiewicz, A.~Fronczak, P.~Fronczak, and K.~Suchecki,
\newblock ``Universal scaling of distances in complex networks,''
\newblock {\em Physical Review E}, vol. 72, pp. 026108, 2005.

\bibitem{AlDo04}
M.~J. Alava and S.~N. Dorogovtsev,
\newblock ``Preferential compactness of networks,''
  \url{arXiv:cond-mat/0407643}.

\bibitem{BeBoBo03}
N.~Berger, B.~Bollob\'{a}s, C.~Borgs, J.~T. Chayes, and O.~Riordan,
\newblock ``Degree distribution of the {FKP} network model,''
\newblock in {\em {ICALP}}, 2003.

\bibitem{DorMen-book03}
S.~N. Dorogovtsev and J.~F.~F. Mendes,
\newblock {\em Evolution of Networks: From Biological Nets to the {Internet}
  and {WWW}},
\newblock Oxford University Press, Oxford, 2003.

\bibitem{aol-data}
M.~Arrington,
\newblock ``{AOL} proudly releases massive amounts of private data,'' 2006,
\newblock \url{http://www.techcrunch.com/2006/08/06/}.

\bibitem{wit}
{CAIDA},
\newblock ``{Workshop on the Internet Topology},'' 2006,
\newblock \url{http://www.caida.org/workshops/isma/0605/}.

\bibitem{DaAlHaBaVaVe05}
L.~Dall'Asta, I.~Alvarez-Hamelin, A.~Barrat, A.~V\'{a}zquez, and A.~Vespignani,
\newblock ``Exploring networks with traceroute-like probes: Theory and
  simulations,''
\newblock {\em Theoretical Computer Science, Special Issue on Complex
  Networks}, 2005.

\bibitem{ViBaDaZhKo05}
F.~Viger, A.~Barrat, L.~Dall'Asta, C.~Zhang, and E.~Kolaczyk,
\newblock ``Network inference from traceroute measurements: {Internet} topology
  `species','' \url{arXiv:cs.NI/0510007}.

\bibitem{ChaJaMa05}
H.~Chang, S.~Jamin, Z.~M. Mao, and W.~Willinger,
\newblock ``An empirical approach to modeling inter-{AS} traffic matrices,''
\newblock in {\em IMC}, 2005.

\bibitem{MaKrFo06}
P.~Mahadevan, D.~Krioukov, M.~Fomenkov, B.~Huffaker, X.~Dimitropoulos,
  kc~claffy, and A.~Vahdat,
\newblock ``The {Internet} {AS}-level topology: Three data sources and one
  definitive metric,''
\newblock {\em Computer Communication Review}, vol. 36, no. 1, 2006.

\bibitem{mms04-book}
S.~Attinger and P.~D.~Koumoutsakos (Edts.),
\newblock {\em Multiscale Modelling and Simulation},
\newblock Springer, Berlin, 2004.

\bibitem{mms-journal}
T.~Y.~Hou (Edt.),
\newblock {\em Multiscale Modeling and Simulation: A {SIAM} Interdisciplinary
  Journal},
\newblock SIAM, Philadelphia.

\bibitem{ZhZhZh06}
S.~Zhou, G.-Q. Zhang, and G.-Q. Zhang,
\newblock ``Chinese {Internet} {AS}-level topology,'' 2006,
\newblock \url{arXiv:cs.NI/0511101}.

\bibitem{pdb}
``{Worldwide Protein Data Bank},'' \url{http://www.wwpdb.org}.

\bibitem{datcat}
{CAIDA},
\newblock ``{Internet Measurement Data Catalog},''
  \url{http://imdc.datcat.org}.

\bibitem{datcat-ccr}
C.~Shannon, D.~Moore, K.~Keys, M.~Fomenkov, B.~Huffaker, and kc~claffy,
\newblock ``{The Internet Measurement Data Catalog},''
\newblock {\em Computer Communication Review}, vol. 35, no. 5, 2005.

\bibitem{geni}
``{The GENI Initiative},'' \url{http://www.nsf.gov/cise/geni/}.

\end{thebibliography}
\end{document}